\documentclass[conference]{IEEEtran}
\IEEEoverridecommandlockouts
\usepackage{cite}
\usepackage{amsmath,amssymb,amsfonts}
\usepackage{algorithmic}
\usepackage{graphicx}
\usepackage{textcomp}
\usepackage{xcolor}
\def\BibTeX{{\rm B\kern-.05em{\sc i\kern-.025em b}\kern-.08em
		T\kern-.1667em\lower.7ex\hbox{E}\kern-.125emX}}
\usepackage{multirow}
\usepackage{booktabs}
\usepackage[hidelinks]{hyperref}
\usepackage{tikz}
\usepackage{placeins}
\renewcommand{\Vec}{\mathbf}
\newcommand\norm[1]{\left\lVert#1\right\rVert}
\DeclareMathOperator*{\argmin}{argmin}
\usepackage{siunitx}
\usepackage{enumitem}

\title{A Comparative Analysis of Generalised Echo and Interference Cancelling and Extended Multichannel Wiener Filtering for Combined Noise Reduction and Acoustic Echo Cancellation
	\thanks{This research was carried out at the ESAT Laboratory of KU Leuven, in the frame of Research Council KU Leuven C14-21-0075 "A holistic approach to the design of integrated and distributed digital signal processing algorithms for audio and speech communication devices", and Aspirant Grant 11PDH24N (for A. Roebben) from the Research Foundation - Flanders (FWO).}
}

\author{\IEEEauthorblockN{Arnout Roebben, Toon van Waterschoot, and Marc Moonen}
	\IEEEauthorblockA{\textit{Department of Electrical Engineering (ESAT)}\\ \textit{STADIUS Center for Dynamical Systems, Signal Processing and Data Analytics}\\
	\textit{KU Leuven}\\
	Leuven, Belgium \\
	\{arnout.roebben, toon.vanwaterschoot, marc.moonen\}@kuleuven.be}}

\begin{document}
\IEEEpubid{\makebox[\columnwidth]{\copyright2025 IEEE \hfill}
\hspace{\columnsep}\makebox[\columnwidth]{ }}
\maketitle
\IEEEpubidadjcol

\begin{abstract}
	Two algorithms for combined acoustic echo cancellation (AEC) and noise reduction (NR) are analysed, namely the generalised echo and interference canceller (GEIC) and the extended multichannel Wiener filter (MWF\textsubscript{ext}). Previously, these algorithms have been examined for linear echo paths, and assuming access to voice activity detectors (VADs) that separately detect desired speech and echo activity. However, algorithms implementing VADs may introduce detection errors. Therefore, in this paper, the previous analyses are extended by 1) modelling general nonlinear echo paths by means of the generalised Bussgang decomposition, and 2) modelling VAD error effects in each specific algorithm, thereby also allowing to model specific VAD assumptions. It is found and verified with simulations that, generally, the MWF\textsubscript{ext} achieves a higher NR performance, while the GEIC achieves a more robust AEC performance.
\end{abstract}


\section{Introduction}
\label{sec:introduction}
In various speech recording applications, next to desired speech, unwanted noise and acoustic echo are recorded. The noise originates from within the room (the near-end room) while the echo originates from recordings in another room (the far-end room) that are played by loudspeakers in the near-end room \cite{hanslerTopicsAcousticEcho2006}. To reduce the interferers, an algorithm for combined acoustic echo cancellation (AEC) and noise reduction (NR) is required \cite{hanslerTopicsAcousticEcho2006,herbordtJointOptimizationLCMV2004,ruizDistributedCombinedAcoustic2022,romboutsIntegratedApproachAcoustic2005a}. Next to algorithms that cascade separate AEC and NR algorithms \cite{schrammenChangePredictionLow2019,cohenJointBeamformingEcho2018}, integrated algorithms can be derived from a single cost function, such as the generalised echo and interference canceller (GEIC) \cite{herbordtJointOptimizationLCMV2004} and the extended multichannel Wiener filter (MWF\textsubscript{ext}) \cite{ruizDistributedCombinedAcoustic2022,romboutsIntegratedApproachAcoustic2005a}.

The GEIC has been analysed in \cite{maruoOptimalSolutionsBeamformer2011a,sprietCombinedFeedbackNoise2007}, and the MWF\textsubscript{ext} in \cite{ruizDistributedCombinedAcoustic2022}. These studies, however, consider only linear echo paths, be it that in \cite{maruoOptimalSolutionsBeamformer2011a} undermodelling is also considered. In addition, the algorithms assume access to voice activity detectors (VADs) that separately detect desired speech and echo activity. However, algorithms implementing VADs may introduce detection errors \cite{zhuRobustLightweightVoice2023a,hamidiaNewRobustDoubletalk2017}. 

In this paper, the previous analyses are extended by 1) modelling general nonlinear echo paths by means of the generalised Bussgang decomposition \cite{demirBussgangDecompositionNonlinear2021}, and 2) modelling VAD error effects in each specific algorithm. This analysis framework also allows to study specific VAD assumptions, e.g., considering a permanent doubletalk scenario where the desired speech and echo are simultaneously active \cite{sprietCombinedFeedbackNoise2007}.  It is found that, generally, the MWF\textsubscript{ext} achieves a higher NR performance, while the GEIC achieves a more robust AEC performance. Coding examples are available \cite{roebbenGithubRepositoryComparative2024}.

\section{Signal model}
\label{sec:signal_model}
The signal model is presented in the $z$-domain. The $z$- to frequency-domain conversion is realised by replacing index $z$ with frequency-bin index $f$ (and possibly frame index $k$). The $z$- to time-domain conversion is realised by replacing the $z$-domain variables with time-lagged vectors. 

An $M$-microphone/$L$-loudspeaker scenario is considered, where $m_i(z)$ represents the microphone signal in microphone $i\!=\!\{1,...,M\}$, and $l_j(z)$ represents the loudspeaker signal in loudspeaker $j\!=\!\{1,...,L\}$. The signals $m_i(z)$ can then be collected in a microphone signal vector $\Vec{m}(z)\in\mathbb{C}^{M\times 1}$ as
\begin{equation} \label{eq:signal_model_def_m_z_domain}
	\Vec{m}(z) = \begin{bmatrix}
		m_1(z)&
		m_2(z)&
		\dots&
		m_M(z)
	\end{bmatrix}^\top_{\textstyle \raisebox{2pt}{,}}
\end{equation}
which consists of a desired speech signal vector $\Vec{s}(z)$, noise signal vector $\Vec{n}(z)$ and echo signal vector $\Vec{e}(z)$ as
\begin{equation}
	\Vec{m}(z) =\Vec{s}(z) + \Vec{n}(z) + \Vec{e}(z)_{\textstyle \raisebox{2pt}{.}}
\end{equation} 
Herein, $\Vec{s}(z)$, $\Vec{n}(z)$ and $\Vec{e}(z)$ are assumed mutually uncorrelated. The echo signal vector $\Vec{e}(z)=F(\Vec{l}(z))$ originates from the loudspeaker signal vector $\Vec{l}(z)\in\mathbb{C}^{L\times 1}$
\begin{equation} \label{eq:signal_model_def_l_z_domain}
	\Vec{l}(z) = \begin{bmatrix}
		l_1(z)&
		l_2(z)&
		\dots&
		l_L(z)
	\end{bmatrix}^\top_{\textstyle \raisebox{2pt}{,}}
\end{equation}
with a map $F(.):\mathbb{C}^{L\times 1}\to\mathbb{C}^{M\times 1}$ representing the echo path from loudspeakers to microphones. Further, an extended microphone signal vector $\tilde{\Vec{m}}(z)\in\mathbb{C}^{(M+L)\times 1}$ is defined by stacking $\Vec{m}(z)$ and $\Vec{l}(z)$ as
\begin{subequations} \label{eq:signal_model_def_x_tilde}
	\begin{align}
		\tilde{\Vec{m}}(z) &= \begin{bmatrix} \Vec{m}(z)^\top & \Vec{l}(z)^\top \end{bmatrix}^\top\\
		&= \tilde{\Vec{s}}(z)+\tilde{\Vec{n}}(z) + \tilde{\Vec{e}}(z)\\
		&= \begin{bmatrix} \Vec{s}(z) \\ \Vec{0}_{L\times1} \end{bmatrix}+\begin{bmatrix} \Vec{n}(z) \\ \Vec{0}_{L\times1} \end{bmatrix}+\begin{bmatrix} \Vec{e}(z) \\ \Vec{l}(z) \end{bmatrix}_{\textstyle \raisebox{2pt}{.}}
	\end{align}
\end{subequations}	
For a single desired speech source, $\Vec{s}(z)$ can be represented as $\Vec{s}(z)=  \Vec{h}(z)s_r(z)$ with $\Vec{h}(z)=\begin{bmatrix}{h_1}(z)/{h_r(z)} &... & {h_M}(z)/{h_r(z)}\end{bmatrix}^\top\in\mathbb{C}^{M\times 1}$ collecting the relative transfer functions $\frac{h_i(z)}{h_r(z)}$ with $h_i(z)$ the transfer function from source $s(z)$ to microphone $i$, and $s_r(z)=  h_r(z)s(z)$ the desired speech in reference microphone $r$. Similarly, $\tilde{\Vec{s}}(z) = \tilde{\Vec{h}}(z)s_r(z)$ with $\tilde{\Vec{h}}(z) = \begin{bmatrix}
	\Vec{h}(z)^\top & \Vec{0}_{L\times 1}^\top
\end{bmatrix}^\top_{\textstyle \raisebox{2pt}{.}}$ From here on, z-indices are omitted for conciseness.

While the noise is assumed to be always-on, the desired speech and echo are on-off with near-end and far-end speaker pauses. This desired speech and echo activity are independent, resulting in four regimes, during which the algorithms estimate the required signal statistics \cite{hanslerTopicsAcousticEcho2006,herbordtJointOptimizationLCMV2004,romboutsIntegratedApproachAcoustic2005a,ruizDistributedCombinedAcoustic2022}. Thus, separate voice activity detectors (VADs) \cite{zhuRobustLightweightVoice2023a,hamidiaNewRobustDoubletalk2017} are required for the desired speech ($\text{VAD}_s$) and echo ($\text{VAD}_e$) to distinguish desired speech activity ($\text{VAD}_s  = 1$) and inactivity ($\text{VAD}_s =  0$), and echo activity ($\text{VAD}_e  =  1$) and inactivity ($\text{VAD}_e  =  0$). 

With $\mathbb{E}_{k}\{.\}$ the expected value in periods where $\Vec{k}\in\{\Vec{s},\Vec{e}\}$ is active (i.e., $\text{VAD}_k =1$), and assuming (short-term) stationarity and ergodicity during the activity (e.g., modelling a batch of data), the desired speech and echo correlation matrix can be computed as $\mathbb{E}_{s}\{\Vec{s}\Vec{s}^H\}= R_{ss}\in\mathbb{C}^{M\times M}$, and $\mathbb{E}_{e}\{\Vec{e}\Vec{e}^H\}=  R_{ee}$ respectively. As the noise is always-on, $\mathbb{E}_{s}\{{\Vec{n}}{\Vec{n}}^H\}=\mathbb{E}_{e}\{{\Vec{n}}{\Vec{n}}^H\}=\mathbb{E}\{{\Vec{n}}{\Vec{n}}^H\}  =  R_{{n}{n}}$. Similarly, the extended desired speech, echo and noise correlation matrix are respectively defined as
\begin{subequations}
	\begin{align}
		&\mathbb{E}_{s}{\{\tilde{s}\tilde{s}^H\}} = R_{\tilde{s}\tilde{s}} = \begin{bmatrix}
			R_{ss}  & 0_{M\times L}\\ 0_{L\times M} & 0_{L\times L}
		\end{bmatrix}\\
		&\mathbb{E}_{e}{\{\tilde{e}\tilde{e}^H\}} = R_{\tilde{e}\tilde{e}} = \begin{bmatrix}
		R_{ee} & R_{el}\\ R_{le} & R_{ll}
	\end{bmatrix}\\
		&\mathbb{E}_{s}{\{\tilde{n}\tilde{n}^H\}} = \mathbb{E}_{e}{\{\tilde{n}\tilde{n}^H\}} = R_{\tilde{n}\tilde{n}} = \begin{bmatrix}
			R_{nn} & 0_{M\times L}\\ 0_{L\times M} & 0_{L\times L}
		\end{bmatrix}_{\textstyle \raisebox{2pt}{,}}
	\end{align}
\end{subequations} 
with $\mathbb{E}_{{e}}\{\Vec{l}\Vec{l}^H\}= R_{ll}\in\mathbb{C}^{L\times L}$ and $\mathbb{E}_{{e}}\{\Vec{e}\Vec{l}^H\}= R_{el}\in\mathbb{C}^{M\times L}$ the loudspeaker and echo-loudspeaker correlation matrix.

However, VAD algorithms may introduce detection errors \cite{zhuRobustLightweightVoice2023a,hamidiaNewRobustDoubletalk2017}, e.g., erroneously detecting desired speech as echo since both signals are speech signals with similar characteristics. Therefore, the (extended) microphone correlation matrices below will be re-defined by considering VAD errors. Furthermore, this framework will allow to study the effect of specific VAD assumptions, e.g., the effect of a (possibly error-free) VAD on the microphone signals only. This VAD is then only able to distinguish between noise-only periods and periods of desired speech or echo activity, or both, as both signals are speech signals with similar characteristics. Similarly, a scenario with permanent doubletalk can be studied, where the desired speech and echo are simultaneously active, such that $\text{VAD}_s$ and $\text{VAD}_e$ coincide \cite{sprietCombinedFeedbackNoise2007}. 

As will be seen in Section \ref{sec:GEIC}, the GEIC requires access to the regime $\{\text{VAD}_s =0 \land \text{VAD}_e =1\}$, corresponding to a correlation matrix $R_{\tilde{n}\tilde{n}}+R_{\tilde{e}\tilde{e}}$. However, with VAD errors, this correlation matrix will be replaced by $R_{\tilde{m}\tilde{m}}^{\{
	\alpha_s,\alpha_e\}}$ with
\begin{equation} \label{eq:Rmm1}
		R_{\tilde{m}\tilde{m}}^{\{
			\alpha_s,\alpha_e\}}=\alpha_s R_{\tilde{s}\tilde{s}}+R_{\tilde{n}\tilde{n}}+ \alpha_e R_{{\tilde{e}\tilde{e}}{\textstyle \raisebox{2pt}{.}}}
\end{equation}
Herein, $\alpha_s,\alpha_e \in[0,1]$ are scaling factors arising when calculating the average correlation matrix. Indeed, due to VAD errors, in some periods the echo (desired speech) will be inactive (active), thereby reducing (increasing) its contribution in the average correlation matrix. Without VAD errors, $\alpha_s =  0$ and $\alpha_e = 1$. Considering an error-free VAD in the permanent doubletalk scenario yields $\alpha_s  = \alpha_e = 1$.

Similarly, it will be seen in Section \ref{sec:MWFext} that the MWF\textsubscript{ext} requires access to the regimes $\{\text{VAD}_s =1 \land \text{VAD}_e =1\}$ and $\{\text{VAD}_s =0 \land \text{VAD}_e =1\}$, which correspond to the correlation matrices $R_{\tilde{s}\tilde{s}}+R_{\tilde{n}\tilde{n}}+R_{\tilde{e}\tilde{e}}$ and $R_{\tilde{n}\tilde{n}}+R_{\tilde{e}\tilde{e}}$ respectively. However, with VAD errors, these correlation matrices will be replaced by $R_{\tilde{m}\tilde{m}}^{\{\beta_s,\beta_e\}}$ and $R_{\tilde{m}\tilde{m}}^{\{\gamma_s,\gamma_e\}}$ respectively with
\begin{subequations} 
	\begin{align}
	&R_{\tilde{m}\tilde{m}}^{\{
		\beta_s, \beta_e\}}=\beta_s R_{\tilde{s}\tilde{s}}+R_{\tilde{n}\tilde{n}}+ \beta_e R_{{\tilde{e}\tilde{e}}}\label{eq:Rmm2}\\
	&R_{\tilde{m}\tilde{m}}^{\{
		\gamma_s, \gamma_e\}}=\gamma_s R_{\tilde{s}\tilde{s}}+R_{\tilde{n}\tilde{n}}+ \gamma_e R_{{\tilde{e}\tilde{e}}{\textstyle \raisebox{2pt}{.}}}\label{eq:Rmm3}
	\end{align}
\end{subequations}	
Herein, $\beta_s, \beta_e, \gamma_s, \gamma_e  \in [0,1]$ are again scaling factors. Without VAD errors, $\beta_s  =  \beta_e  =  \gamma_e  =  1$ and $\gamma_s  =  0$. Considering an (error-free) VAD only operating on the microphone signals, which is then only able to distinghuish between noise-only periods and periods of desired speech or echo activity, or both, $\gamma_s  = \gamma_e = 0$, and considering an (error-free) VAD in the permanent doubletalk scenario, $\beta_s  =  \beta_e  = 1$ and $\gamma_s  = \gamma_e = 0$.

$R_{ss}$ is assumed rank $1$, modelling a single desired speaker, and $R_{nn}$ and $R_{ll}$ are assumed full rank.

\section{Filter design} 
\subsection{Generalised Bussgang decomposition} \label{sec:Generalised_Bussgang_decomposition}
To analyse a general nonlinear echo path map $F(.)$, the generalised Bussgang decomposition is applied to decompose the echo signal vector $\Vec{e}$ into two uncorrelated components, $\Vec{e}_{\text{lin}}\in\mathbb{C}^{M\times 1}$ and $\Vec{e}_{\text{res}}\in\mathbb{C}^{M\times 1}$ \cite{demirBussgangDecompositionNonlinear2021}. Herein, $\Vec{e}_{\text{lin}}=F_{\text{lin}}^H\Vec{l}$ represents the linear echo component, with $F_{\text{lin}}=R_{ll}^{-1}R_{le}$ the minimum mean squared error optimal estimate of the echo path \cite{romboutsAdaptiveFilteringAlgorithms2003}, and $\Vec{e}_{\text{res}}=\Vec{e}-F_{\text{lin}}^H\Vec{l}$ represents the residual echo component after subtracting $\Vec{e}_{\text{lin}}$ from $\Vec{e}$. Indeed, $\Vec{e}_{\text{lin}}$ and $\Vec{e}_{\text{res}}$ are uncorrelated as \cite{demirBussgangDecompositionNonlinear2021}
\begin{equation}
	\begin{aligned}
		\mathbb{E}\left\{\Vec{e}_{\text{res}}\Vec{l}^H\right\}=0_{M\times L} \quad \land \quad
		\mathbb{E}\left\{\Vec{e}_{\text{res}}\Vec{e}_{\text{lin}}^H\right\}=0_{M\times M_{\textstyle \raisebox{2pt}{.}}}
	\end{aligned}
\end{equation}
Therefore, $R_{\tilde{e}\tilde{e}}$ can be split into two uncorrelated components:
\begin{subequations}
	\begin{align}
		R_{\tilde{e}\tilde{e}} &= R_{\tilde{e}_{\text{lin}}\tilde{e}_{\text{lin}}} + R_{\tilde{e}_{\text{res}}\tilde{e}_{\text{res}}}\\&=\begin{bmatrix}
			F_{\text{lin}}^HR_{ll}F_{\text{lin}} & F_{\text{lin}}^HR_{ll}\\ R_{ll}F_{\text{lin}} & R_{ll}
		\end{bmatrix} + \begin{bmatrix}
			R_{e_{\text{res}}e_{\text{res}}} & 0_{M\times L}\\ 0_{L\times M} & 0_{L\times L}
		\end{bmatrix}_{\textstyle \raisebox{2pt}{.}} \label{eq:def_Rmm}
	\end{align}
\end{subequations}

\subsection{Generalised echo and interference canceller (GEIC)} \label{sec:GEIC}
The GEIC \cite{herbordtJointOptimizationLCMV2004} aims at designing a filter $\tilde{\Vec{w}}_{\text{GEIC}}\in\mathbb{C}^{(M+L)\times 1}$ that minimises the filter output power while preserving the desired speech in the reference microphone
\begin{equation}
	\begin{aligned}
		\tilde{\Vec{w}}_{\text{GEIC}}=\argmin_{\tilde{\Vec{w}}} \quad & \tilde{\Vec{w}}^HR_{\tilde{m}\tilde{m}}^{\{\alpha_s,\alpha_e\}}\tilde{\Vec{w}}\\
		\textrm{s.t.} \quad & \tilde{\Vec{w}}^H\tilde{\Vec{h}} = 1_{\textstyle \raisebox{2pt}{,}}\\
	\end{aligned}
\end{equation}
of which the solution equals \cite{herbordtJointOptimizationLCMV2004}:
\begin{equation}
	\tilde{\Vec{w}}_{\text{GEIC}} = \begin{bmatrix}
		\Vec{w}_c - B\Vec{w}_a\\-\Vec{a}
	\end{bmatrix}_{\textstyle \raisebox{2pt}{.}}
\end{equation}
Herein, a quiescent beamformer $\Vec{w}_c = \frac{\Vec{h}}{\Vec{h}^H\Vec{h}}\in\mathbb{C}^{M\times 1}$ is defined, as well as a blocking matrix $B\in\mathbb{C}^{M\times (M-1)}$ with $\Vec{h}^HB=\Vec{0}_{(M-1)\times 1}^\top$, and $\Vec{w}_a\in\mathbb{C}^{(M-1) \times 1}$ and $\Vec{a}\in\mathbb{C}^{L\times 1}$ are jointly optimised by means of an adaptive filter procedure, which, using \cite[(2.3)]{luInversesBlockMatrices2002}, converges to \cite{herbordtJointOptimizationLCMV2004,maruoOptimalSolutionsBeamformer2011a}
\begin{equation}\label{eq:GEIC_wa_a2}
	\begin{bmatrix}
		\Vec{w}_a \\ \Vec{a}
	\end{bmatrix}=\begin{bmatrix}
	S^{-1}(B^HR_{mm}^{\{\alpha_s,\alpha_e\}}\Vec{w}_c- \alpha_e B^HR_{el}R_{ll}^{-1}R_{le}\Vec{w}_c)\\
	-R_{ll}^{-1}R_{le}B\Vec{w}_a+R_{ll}^{-1}R_{le}\Vec{w}_{c}
\end{bmatrix}_{\textstyle \raisebox{2pt}{,}}
\end{equation}
with $S = B^HR_{mm}^{\{\alpha_s,\alpha_e\}}B- \alpha_e B^HR_{el}R_{ll}^{-1}R_{le}B$. Referring to the generalised Bussgang decomposition (Section \ref{sec:Generalised_Bussgang_decomposition}), $\Vec{a}$ aims at modelling the linear component of the overall echo path from the loudspeakers to the output of the GEIC, consisting of a combination of the optimal mean squared error estimate of the echo path $F_{\text{lin}}=R_{ll}^{-1}R_{le}$, with the parallel filters $\Vec{w}_c$ and $-B\Vec{w}_a$. Thus, $\Vec{a}$ allows for a perfect cancellation of $\Vec{e}_{\text{lin}}$ regardless of $\alpha_e$, i.e., $\tilde{\Vec{w}}_{\text{GEIC}}^H
\tilde{\Vec{e}}_{\text{lin}}=\tilde{\Vec{w}}_{\text{GEIC}}^H\begin{bmatrix}\Vec{l}^HF_\text{lin} & \Vec{l}^H\end{bmatrix}^H=0$. On the other hand, $\Vec{w}_a$ aims at estimating the noise and residual echo after applying $\Vec{w}_c$, from the noise and residual echo after applying $B$. Indeed, $\Vec{w}_a$ (\ref{eq:GEIC_wa_a2}) can be simplified as
\begin{equation}
	\small\begin{aligned}
	\Vec{w}_a \!= \!(B^H(R_{nn}\!+\! \alpha_e R_{e_{\text{res}}e_{\text{res}}})B)^{{-1}}(B^H(R_{nn}\!+\! \alpha_e R_{e_{\text{res}}e_{\text{res}}})\Vec{w}_c)_{\textstyle \raisebox{2pt}{,}}
	\end{aligned}
\end{equation}
such that the $\Vec{w}_a$ reduces noise and residual echo. When the echo is linear, i.e., $\Vec{e}_{\text{res}}=\Vec{0}_{M\times 1}$, $\Vec{a}$ removes the entire echo, and $\Vec{w}_a$ solely reduces the noise, which is consistent with \cite{sprietCombinedFeedbackNoise2007}. 

An imperfect blocking matrix ($\Vec{h}^HB\neq\Vec{0}_{(M-1)\times 1}^\top$) leads to suppression of the desired speech, such that $\Vec{w}_a$ should be updated ideally under desired speech inactivity and echo activity. Indeed, as $\alpha_s$ decreases due to fewer VAD errors, less weight is put on reducing the desired speech.

\subsection{Extended multichannel Wiener filter (MWF\textsubscript{ext})} \label{sec:MWFext}
The MWF\textsubscript{ext} as proposed by \cite{romboutsIntegratedApproachAcoustic2005a,ruizDistributedCombinedAcoustic2022} aims at designing a filter $\tilde{\Vec{w}}_{\text{MWF\textsubscript{ext}}}\in\mathbb{C}^{(M+L)\times 1}$ to retain the desired speech in the reference microphone $r$, while reducing noise and echo:
\begin{equation} \label{eq:MWFext_cost}
		\tilde{\Vec{w}}_{\text{MWF\textsubscript{ext}}} = \argmin_{\tilde{\Vec{w}}} \mathbb{E}\left\{\norm{s_r-\tilde{\Vec{w}}^H\tilde{\Vec{m}}}^2_2\right\}_{\textstyle \raisebox{2pt}{,}}
\end{equation}
which leads to:
\begin{equation} \label{eq:MWFext}
	\tilde{\Vec{w}}_{\text{MWF\textsubscript{ext}}} = R_{\tilde{m}\tilde{m}}^{-1}R_{\tilde{s}\tilde{s}}\tilde{\Vec{t}}_{r_{\textstyle \raisebox{2pt}{,}}}
\end{equation}
where $\tilde{\Vec{t}}_r\in\mathbb{C}^{(M+L)\times 1}$ corresponds to a unit vector with all zeros except at position $r$. Without VAD errors, $R_{\tilde{s}\tilde{s}}$ can be computed by subtracting the correlation matrix recorded when $\{\text{VAD}_s =0 \land \text{VAD}_e =1\}$ from the correlation matrix recorded when $\{\text{VAD}_s =1 \land \text{VAD}_e =1\}$ \cite{benestySpeechEnhancement2005}. However, with VAD errors, this subtraction of (\ref{eq:Rmm3}) from (\ref{eq:Rmm2}) yields $(\beta_s - \gamma_s) R_{\tilde{s}\tilde{s}}+(\beta_e - \gamma_e) R_{\tilde{e}_{\text{lin}}\tilde{e}_{\text{lin}}}+(\beta_e - \gamma_e) R_{\tilde{e}_{\text{res}}\tilde{e}_{\text{res}}}$ rather than $R_{\tilde{s}\tilde{s}}$.

It is shown that using a generalised eigenvalue decomposition (GEVD), remarkably, the linear echo component can be removed entirely despite it being observed together with the desired speech, while the residual echo cannot be removed entirely. Thereby the GEVD procedure of \cite{ruizDistributedCombinedAcoustic2022} proposed for linear echo paths is extended to general nonlinear echo paths. The GEVD of $\{R_{\tilde{m}\tilde{m}}^{\{\beta_s,\beta_e\}},R_{\tilde{m}\tilde{m}}^{\{\gamma_s,\gamma_e\}}\}$ is defined as
\begin{subequations}
	\begin{align}
		R_{\tilde{m}\tilde{m}}^{\{\beta_s,\beta_e\}} &=Q\text{diag}(\lambda_{\tilde{m}_1}^{\{\beta_s,\beta_e\}},...,\lambda^{\{\beta_s,\beta_e\}}_{\tilde{m}_{(M+L)}})Q^H\\
		R_{\tilde{m}\tilde{m}}^{\{\gamma_s,\gamma_e\}} &=Q\text{diag}(\lambda_{\tilde{m}_1}^{\{\gamma_s,\gamma_e\}},...,\lambda^{\{\gamma_s,\gamma_e\}}_{\tilde{m}_{(M+L)}})Q^H_{\textstyle \raisebox{2pt}{,}}
	\end{align}
\end{subequations}
with $Q\in\mathbb{C}^{(M+L)\times(M+L)}$ collecting the generalised eigenvectors in its columns, $\lambda_{\tilde{m}_p}^{\{\beta_s,\beta_e\}}$ and $\lambda_{\tilde{m}_p}^{\{\gamma_s,\gamma_e\}}$ ($p =\{1,..., M + L\}$) defining the generalised eigenvalues, and diag(.) representing the operation to create a diagonal matrix. $R_{\tilde{s}\tilde{s}}$, $R_{\tilde{n}\tilde{n}}$ and $R_{\tilde{e}_{\text{res}}\tilde{e}_{\text{res}}}$ contain a zero structure related to the loudspeaker signals, since $\Vec{s}$, $\Vec{n}$ and $\Vec{e}_{\text{res}}$ are uncorrelated with $\Vec{l}$. However, $R_{\tilde{e}_{\text{lin}}\tilde{e}_{\text{lin}}}$ lacks this zero structure due to the correlation of $\Vec{e}^{lin}$ with $\Vec{l}$. Consequently, $Q$ reflects this structure:
\begin{equation}
	Q =\left[ \begin{array}{c|c@{}}
		Q_1 & \multirow{ 2}{*}{$Q_2$}\\
		\cmidrule{1-1}
		 0_{L\times R}&\\
	\end{array} \right]_{\textstyle \raisebox{2pt}{,}}
\end{equation}
with $R=\text{rank}\!\left((\beta_s\!-\!\gamma_s) R_{\tilde{s}\tilde{s}}+(\beta_e\!-\!\gamma_e) R_{\tilde{e}_{\text{res}}\tilde{e}_{\text{res}}}\right)$. If $R=1$, $\begin{bmatrix} Q_{1}^\top & \Vec{0}_{L\times 1}^\top\end{bmatrix}^\top$ models $\tilde{\Vec{s}}$. If $R=M$, $\begin{bmatrix} Q_{1}^\top & 0_{L\times R}^\top\end{bmatrix}^\top$ is related to $\tilde{\Vec{s}}$, $\tilde{\Vec{n}}$, and $\tilde{\Vec{e}}_{\text{res}}$, and $Q_2$ is related to $\tilde{\Vec{e}}_{\text{lin}}$. Explicitly setting the generalised eigenvalues related to $Q_2$, and thus related to $R_{\tilde{e}_{\text{lin}}\tilde{e}_{\text{lin}}}$, to $0$, the component corresponding to $\tilde{\Vec{e}}_{\text{lin}}$ can still be removed regardless of $\beta_e$ and $\gamma_e$. The estimate for the desired speech correlation matrix $\hat{R}_{\tilde{s}\tilde{s}}$ is then computed as
\begin{equation}
\small	\begin{aligned}
	&\hat{R}_{\tilde{s}\tilde{s}}\!=\! Q\text{diag}({\lambda}^{\{\beta_s,\beta_e\}}_{\tilde{m}_1}\!\!-\!{\lambda}^{\{\gamma_s,\gamma_e\}}_{\tilde{m}_1}\!\!,\!...,{\lambda}^{\{\beta_s,\beta_e\}}_{\tilde{m}_{M}}\!\!-\!{\lambda}^{\{\gamma_s,\gamma_e\}}_{\tilde{m}_{M}},0,\!...,0)Q^H\label{eq:GEVD1}\\
&= (\beta_s-\gamma_s) R_{\tilde{s}\tilde{s}} + (\beta_e-\gamma_e) R_{\tilde{e}_{\text{res}}\tilde{e}_{\text{res}}{\textstyle \raisebox{2pt}{.}}}
	\end{aligned}
\end{equation}
Plugging $\hat{R}_{\tilde{s}\tilde{s}}$ into (\ref{eq:MWFext}) leads to the following realisable filter:
\begin{equation}\label{eq:MWFext2}
\hat{\tilde{\Vec{w}}}_{\text{MWF\textsubscript{ext}}}= R_{\tilde{m}\tilde{m}}^{-1}\!\left((\beta_s\!-\!\gamma_s) R_{\tilde{s}\tilde{s}}+(\beta_e\!-\!\gamma_e) R_{\tilde{e}_{\text{res}}\tilde{e}_{\text{res}}}\right)\tilde{\Vec{t}}_{r_{\textstyle \raisebox{2pt}{,}}}
\end{equation}
thereby reducing the noise and the linear echo component, but not the residual echo. Indeed, a weighted average of the desired speech and residual echo is being reconstructed, and $\hat{R}_{\tilde{s}\tilde{s}}$ can even be negative definite due to the VAD errors. Explicitly setting the rank of $\hat{R}_{\tilde{s}\tilde{s}}$ to its theoretical value of $1$ is realised by sorting the ratio of generalised eigenvalues,
\begin{equation} \label{eq:sorted_eigenvalues}
	\begin{aligned}
		\bigl\{{\lambda}^{\{\beta_s,\beta_e\}}_{\tilde{m}_1}/{\lambda}^{\{\gamma_s,\gamma_e\}}_{\tilde{m}_1},...,{\lambda}^{\{\beta_s,\beta_e\}}_{\tilde{m}_M}/{\lambda}^{\{\gamma_s,\gamma_e\}}_{\tilde{m}_M} \bigr\}_{\textstyle \raisebox{2pt}{,}}
	\end{aligned}
\end{equation}
according to an increased ratio and only keeping this largest ratio \cite{ruizDistributedCombinedAcoustic2022,serizelLowrankApproximationBased2014}. Although this further reduces the residual echo as eigenvalue modes containing residual echo are removed, partial desired speech cancellation or residual echo reconstruction might occur. Indeed, the matrix pencil $\{R_{\tilde{m}\tilde{m}}^{\{\beta_s,\beta_e\}},R_{\tilde{m}\tilde{m}}^{\{\gamma_s,\gamma_e\}}\}$ is jointly diagonalised, but $R_{\tilde{s}\tilde{s}}$ not necessarily. 

When the spatial structure of $R_{\tilde{e}_\text{lin}\tilde{e}_\text{lin}}$ would differ between both sides of $\{R_{\tilde{m}\tilde{m}}^{\{\beta_s,\beta_e\}},R_{\tilde{m}\tilde{m}}^{\{\gamma_s,\gamma_e\}}\}$, e.g., due to changing echo paths, the zero-structure in $Q$ would however be destroyed. 

\subsection{Relation between GEIC and MWF\textsubscript{ext}} \label{sec:Relation_GEIC_MWFext}
When $R_{\tilde{s}\tilde{s}}$ is rank-$1$, (\ref{eq:MWFext2}) can be decomposed using the ground-truth $\tilde{\Vec{h}}$ according to a similar proof as in \cite{docloAcousticBeamformingHearing2010} as 
\begin{equation}\label{eq:relation_GEIC_MWFext}
	\begin{aligned}
	\hat{\tilde{\Vec{w}}}_{\text{MWF\textsubscript{ext}}}=&\tilde{\Vec{w}}_{\text{GEIC}}^{\{\beta_s,\beta_e\}}\frac{(\beta_s-\gamma_s) P_{y_{\tilde{s}}y_{\tilde{s}}}}{\beta_s P_{y_{\tilde{s}}y_{\tilde{s}}}+P_{y_{\tilde{n}}y_{\tilde{n}}}+\beta_e P_{y_{\tilde{e}_{\text{res}}}y_{\tilde{e}_{\text{res}}}}}\\&+R_{\tilde{m}\tilde{m}}^{-1}\left((\beta_e-\gamma_e) R_{\tilde{e}_{\text{res}}\tilde{e}_{\text{res}}}\right)\tilde{\Vec{t}}_{r{\textstyle \raisebox{2pt}{.}}}
	\end{aligned}
\end{equation}
Herein, $\tilde{\Vec{w}}_{\text{GEIC}}^{\{\beta_s,\beta_e\}}$ is a GEIC computed using (\ref{eq:Rmm2}), $P_{y_{\tilde{k}}y_{\tilde{k}}}\!=\!\tilde{\Vec{w}}_{\text{GEIC}}^{{\{\beta_s,\beta_e\}}^H}\!\!R_{\tilde{k}\tilde{k}}\tilde{\Vec{w}}_{\text{GEIC}}^{{\{\beta_s,\beta_e\}}}\!$ with $\tilde{\Vec{k}}\!\in\!\{\tilde{\Vec{s}},\tilde{\Vec{n}},\tilde{\Vec{e}}_{\text{res}},\tilde{\Vec{e}}_{\text{lin}}\}$, and $P_{y_{\tilde{e}_{\text{lin}}}y_{\tilde{e}_{\text{lin}}}}\!\!\!=\!0$. Thus, the MWF\textsubscript{ext} can be interpreted as a cascade of a GEIC and a single-channel postfilter, parallel to a filter providing the minimum mean squared error optimal residual echo estimate.

Applying the rank-$1$ approximation to $\hat{R}_{\tilde{s}\tilde{s}}$ (Section \ref{sec:MWFext}), i.e., $\hat{R}_{\tilde{s}\tilde{s}} \!=\!\! \left(\!{\lambda}^{\{\beta_s,\beta_e\}}_{\tilde{m}_1}\!\!-\!{\lambda}^{\{\gamma_s,\gamma_e\}}_{\tilde{m}_1}\!\right)\tilde{\Vec{q}}\tilde{\Vec{q}}^H$ with $\tilde{\Vec{q}}$ the generalised eigenvector in $Q$ corresponding to the largest ratio in (\ref{eq:sorted_eigenvalues}), corresponds to replacing the ground-truth $\tilde{\Vec{h}}$ by its estimate $\hat{\tilde{\Vec{h}}}_{\text{\tiny{GEVD}}}=\tilde{\Vec{q}}/\tilde{\Vec{q}}(1) \in \mathbb{C}^{(M+L)\times 1}$, in which the last $L$ elements are zero (Section \ref{sec:MWFext}) \cite{markovich-golanPerformanceAnalysisCovariance2015}. Estimating the correlation matrix of the signal to suppress as $\hat{R}_{\tilde{i}\tilde{i}}=R_{\tilde{m}\tilde{m}}^{\{\beta_s,\beta_e\}}\!-\!\hat{R}_{\tilde{s}\tilde{s}}$, (\ref{eq:MWFext2}), can be decomposed using a similar proof as in \cite{docloAcousticBeamformingHearing2010} as,
\begin{equation} \label{eq:relation_GEIC_MWFext2}
	\hat{\tilde{\Vec{w}}}_{\text{MWF\textsubscript{ext}}}=\tilde{\Vec{w}}_{\text{GEIC,GEVD}}^{\{\beta_s,\beta_e\}}\frac{P_{y_{\tilde{s}}y_{\tilde{s}},\text{\tiny{GEVD}}}}{P_{y_{\tilde{s}}y_{\tilde{s}},\text{\tiny{GEVD}}}+P_{y_{\tilde{i}}y_{\tilde{i}},\text{\tiny{GEVD}}}}{\textstyle \raisebox{2pt}{.}}
\end{equation}
Herein, $\tilde{\Vec{w}}_{\text{GEIC,GEVD}}^{\{\beta_s,\beta_e\}}$ is a GEIC computed using (\ref{eq:Rmm2})  and $\hat{\tilde{\Vec{h}}}_{\text{\tiny{GEVD}}}$, and $P_{y_{\tilde{k}}y_{\tilde{k}},\text{\tiny{GEVD}}}=\tilde{\Vec{w}}_{\text{GEIC,GEVD}}^{{\{\beta_s,\beta_e\}}^H}\hat{R}_{\tilde{k}\tilde{k}}\tilde{\Vec{w}}_{\text{GEIC,GEVD}}^{{\{\beta_s,\beta_e\}}}$ with  $\tilde{\Vec{k}}\in\{\tilde{\Vec{s}},\tilde{\Vec{i}}\}$ determining the succeeding single-channel postfilter.

\section{Simulations} \label{sec:experimental_procedures}
Five scenarios in a $\SI{5}{\meter}$ $\times$ $\SI{5}{\meter}$ $\times$ $\SI{3}{\meter}$ room with $M=2$ microphones at $\begin{bmatrix} 2 & 1.9 & 1\end{bmatrix}$$\SI{}{\meter}$ and $\begin{bmatrix} 2 & 1.8 & 1 \end{bmatrix}$$\SI{}{\meter}$, $L=2$ loudspeakers, $1$ babble noise source \cite{auditecAuditoryTestsRevised1997}, and $1$ desired speech source are studied, with varying positions for the desired speech source, the loudspeakers and the noise source. The desired speech originates from sentences of the Hearing in Noise Test (HINT) database concatenated with $\SI{5}{s}$ of silence \cite{nilssonDevelopmentHearingNoise1994}. The loudspeaker signals contain a mixture of HINT sentences and white noise at a power ratio of $\SI{0}{\decibel}$. All signals are $\SI{10}{\second}$ long. Impulse responses of length $128$ samples at a sampling rate of $\SI{16}{\kilo\hertz}$ are generated using the randomised image method with reflection coefficient $0.15$ and randomised distances of $\SI{0.13}{\meter}$ \cite{desenaModelingRectangularGeometries2015}. The input signal-to-noise ratio (SNR) and signal-to-echo ratio are set to $\SI{5}{\decibel}$. The echo path is modelled either as a linear path characterised by its impulse response, or as a Hammerstein model, i.e., a static nonlinearity $(.)^3$ followed by a linear path. To illustrate the analysis framework, a permanent doubletalk scenario with error-free VAD is considered with $\alpha_s \!= \! \alpha_e  \!= \! \beta_s \! =  \!\beta_e \! =\!  1$ and $\gamma_s \!= \! \gamma_e \! =\! 0$.

The filters are implemented in the short-time Fourier- (STFT) domain using a square-root Hann window and window shift of $512$ and $256$ samples respectively. Correlation matrices are adapted using exponential averaging with a weight for the previous estimate of $0.995$. The rank of $\hat{R}_{\tilde{s}\tilde{s}}$ is set to $1$ using the GEVD procedure. Three GEIC variations are studied: one with access to $\Vec{h}$ (denoted by GEIC), one with access only to the angle of arrival, yielding a delay-and-sum beamformer and imperfect Griffiths-Jim blocking matrix (denoted by GEIC\textsubscript{GJ}) \cite{griffithsAlternativeApproachLinearly1982}, and one using $\hat{\tilde{\Vec{h}}}_{\text{\tiny{GEVD}}}$ (denoted by GEIC\textsubscript{GEVD}).

Performance is evaluated using the intelligibility-weighted SNR improvement $\Delta\text{SNR\textsuperscript{I}}$ (the higher the better), echo return loss enhancement $\text{ERLE\textsuperscript{I}}$ (the higher the better), and speech distortion (SD) $\text{SD\textsuperscript{I}}$ (the closer to zero the better) \cite{greenbergIntelligibilityweightedMeasuresSpeechtointerference1993,makinoSSBSubbandEcho1996,sprietAdaptiveFilteringTechniques2004}. 

\section{Results and discussion} \label{sec:results_and_discussion}
\begin{figure}
		\centering
	\includegraphics[width=0.42\textwidth]{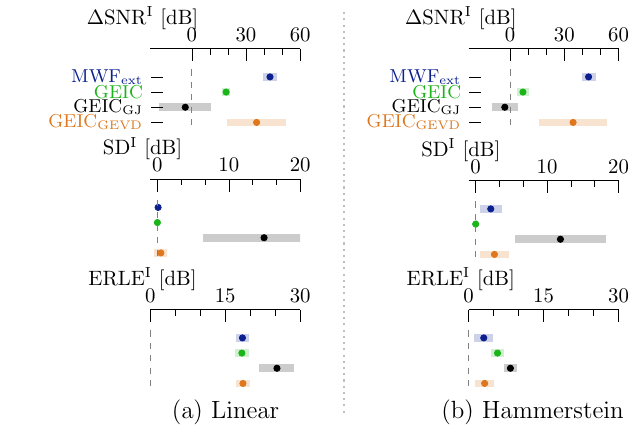}
	\caption{Performance comparison, where dots show the mean performance across scenarios, and shading the standard deviation. The MWF\textsubscript{ext} achieves higher NR, while the GEIC appears more robust regarding the residual echo.}
	\label{fig:res}\end{figure}

While the performance comparison depends on the rank approximation used in (\ref{eq:MWFext2}) and the choice of estimate for $\tilde{\Vec{h}}$, the MWF\textsubscript{ext} generally achieves a higher NR (Fig. \ref{fig:res}) than the GEIC for both echo path types due to the equivalence of (\ref{eq:relation_GEIC_MWFext}) and (\ref{eq:relation_GEIC_MWFext2}), where the presence of the postfilter further reduces the noise with respect to the GEIC. This improved noise reduction comes at the expense of a slightly increased SD\textsuperscript{I} when the GEIC is implemented with a perfect blocking matrix. However, for imperfect blocking matrices, desired speech cancellation occurs, thereby also skewing the $\Delta \text{SNR\textsuperscript{I}}$. 

Regarding the AEC performance, considering a linear echo path, both the GEIC and MWF\textsubscript{ext} achieve similar performance as both similarly suppress linear echo (Section \ref{sec:GEIC} and \ref{sec:MWFext}). However, considering the non-linear echo path, the GEIC appears more robust (Section \ref{sec:MWFext}), although the GEIC assumes access to $\Vec{h}$, which might be unavailable in practice. 

\section{Conclusion} \label{sec:conclusion}
The generalised echo and interference canceller (GEIC) and the extended multichanncel Wiener filter (MWF\textsubscript{ext}) for combined noise reduction (NR) and acoustic echo cancellation (AEC) have been analysed by 1) modelling general nonlinear echo paths using the generalised Bussgang decomposition, and 2) modelling voice activity detection (VAD) error effects in each specific algorithm. Both algorithms suppress the linear echo component regardless of the VAD errors. Generally, the GEIC is found to be more robust towards the residual echo, while the MWF\textsubscript{ext} achieves the highest NR performance. 

\bibliographystyle{IEEEtran}
\bibliography{ref.bib}

\end{document}